\documentclass{jpsj3}
\usepackage{algorithmic}
\usepackage{algorithm}
\usepackage{txfonts}
\usepackage{graphicx}
\usepackage{here}
\usepackage{braket}
\usepackage{bm}
\usepackage{color}   
\usepackage{ulem}
\usepackage[OT1]{fontenc}

\DeclareMathOperator*{\argmax}{arg\,max}

\title{Bayesian Inference for Small-Angle Scattering Data}

\author{Yui Hayashi$^1$, Shun Katakami$^1$, Shigeo Kuwamoto$^2$, Kenji Nagata$^3$,\\ Masaichiro Mizumaki$^2$, and Masato Okada$^1$\thanks{okada@edu.k.u-tokyo.ac.jp}}
\inst{\it{$^1$Graduate School of Frontier Sciences, The University of Tokyo, Kashiwa, Chiba 277-8561, Japan} \\
$^2$Japan Synchrotron Radiation Research Institute (JASRI), Sayo, Hyogo 679-5198, Japan \\ 
$^3$Research and Services Division of Materials Data and Integrated System, National Institute for Materials Science, Tsukuba, Ibaraki 305-0047, Japan} 

\abst{In this paper, we propose a method of estimating model parameters using small-angle scattering (SAS) data based on the Bayesian inference. 
Conventional SAS data analyses involve processes of manual parameter adjustment by analysts or optimization using gradient methods. 
These analysis processes tend to involve heuristic approaches and may lead to local solutions.
Furthermore, it is difficult to evaluate the reliability of the results obtained by conventional analysis methods. 
Our method solves these problems by estimating model parameters as probability distributions from SAS data using the framework of the Bayesian inference. 
We evaluate the performance of our method through numerical experiments using artificial data generated using representative measurement target models.
From the results of numerical experiments, we show that our method provides not only highly accurate and reliable estimations, but also perspectives on the transition point of estimability with respect to the measurement time and the lower bound of the angular region of the measured data.}

\begin{document}
\maketitle

\section{Introduction}
Small-angle scattering (SAS) is a nondestructive measurement method for analyzing nanoscale structural information such as the periodic structure and size of nanoparticles in a measurement target by examining the scattering intensity in a small-angle region among the scatterings produced by irradiating the measurement target with X-rays or neutron beams\cite{saxs1995}. 
The main analysis of SAS data involves fitting a theoretical model to the measured data of scattering vector and scattering intensity to obtain model parameters related to the size and structure of the measurement target. 
Conventionally, these analyses require manual parameter adjustment or nonlinear least-squares fitting using the gradient method. 
These fitting processes are heuristic and lead to local solutions.
Furthermore, it is difficult to evaluate the reliability of the result. \par
In this paper, we solve these problems by estimating the model parameters as probability distributions using the Bayesian inference\cite{hansen2000, asahara2021}. 
 
Our proposed method directly estimates model parameters from measured data, while conventional Bayesian analysis of SAS data, such as that done in BayesApp\cite{BayesApp}, 
performs Bayesian estimation of hyperparameters required to obtain the distance distribution function through indirect Fourier transformation.
In such conventional Bayesian analysis, the scattering intensity can be calculated from the obtained distance distribution function, and the values of the model parameters can be derived, but the model parameters are not directly estimated as probability distributions.
 \par
The probability distributions of model parameters can be calculated from Bayes' theorem as the posterior probability.
The shape and statistics of these probability distributions can be used to obtain the confidence level of estimates. 
The proposed method involves Bayesian spectral analysis \cite{nagata2012, nagata2019} proposed by Nagata {\it et al}., which uses the replica exchange Monte Carlo (EMC) method \cite{hukushima1995}. 
The EMC method is expected to provide globally optimal solutions. 
We describe numerical experiments using monodisperse and polydisperse spheres with Gaussian size distribution, which are typical simple measurement target models in SAS experiments, to evaluate the performance of the proposed method. 
 
From the discussion of the results, we confirm that the proposed method provides not only high accuracy and reliability of estimation,
but also perspectives on the transition point of estimability with respect to the measurement time and the lower bound of the angular region of the measured data.
We also confirmed that we can interpret the change in the reliability of estimation results that occurred due to the decrease in the amount of data.

The proposed method is not limited to the models and parameters treated in these numerical experiments, but can be applied to a wide variety of measurement targets.\par
This paper is organized as follows.
In Sect. 2, we formulate two models of measurement targets, monodisperse and polydisperse spheres with Gaussian size distribution, to be used in numerical experiments.
In Sect. 3, we describe the proposed Bayesian inference method. 
In Sect. 4, we describes numerical experiments on artificial data of various measurement times, angular regions, and extremely small number of data generated using the monodisperse and polydisperse sphere models.  
Then in Sect. 5, we discuss the results of numerical experiments on the reliability of estimates and the transition point of estimability with respect to the measurement time and the lower bound of the angular region of the measured data.
Finally, in Sect. 6, we conclude this paper.

\section{Formulation of Data Generation Process}
In this section, we describe the process of generating SAS data. 
Here, we consider two types of model, monodisperse and polydisperse spheres with Gaussian size distribution, and formulate the scattering intensity and itsmeasured values obtained using these models in Sects. 2.1 and 2.2, respectively\cite{saxs1995,hashimoto-saxs,igor_document}.

\subsection{Monodisperse Spheres}
Here, we formulate the scattering intensity of monodisperse spherical particles with constant radius.
Let $\bm{q}$ be the scattering vector. 
Let $\bm{e}_1$ and $\bm{e}_2$ be the unit vectors in the direction of the wave number vectors of the incident beam   and the scattered beam  , respectively.
The scattering vector $\bm{q}$ is defined as
\begin{equation}
    \bm{q} \equiv \frac{4 \pi \sin \theta}{\lambda} (\bm{e}_2 - \bm{e}_1),
    \label{wave_num}
\end{equation}
where $\lambda$ is the wavelength and the angle $2\theta$ is formed by $\bm{e}_1$ and $\bm{e}_2$ (Fig.1). 

\begin{figure}[H]
    \centering
    \includegraphics[keepaspectratio, scale=0.2]{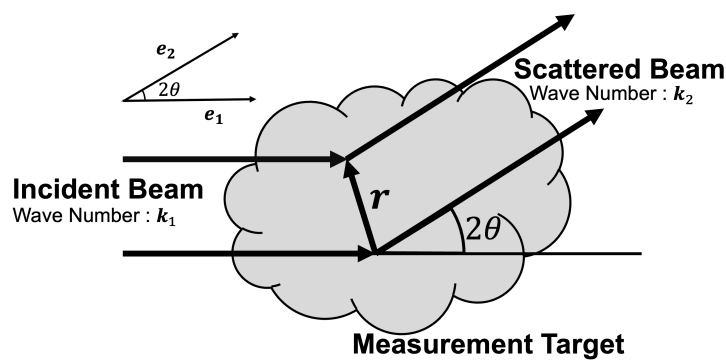}
    \caption{ The positional relationship between the incident and scattered beams.
    The $\bm{e}_1$ and $\bm{e}_2$ are unit vectors in the direction of the $\bm{k}_1$ and $\bm{k}_2$, respectively. }
\end{figure}

\noindent Denoting  the scattering length density   of   scattering elements   at the position vector   in the scatterer   $\bm{r}$ by $\rho(\bm{r})$, the structure factor $F(\bm{q})$ is given by
\begin{equation}
    F(\bm{q}) = \int \rho(\bm{r}) \exp(-i \bm{q} \cdot \bm{r})d\bm{r}.
    \label{F_def}
\end{equation}
Note that $i$ is an imaginary unit. 
In the case of scattering by spherically symmetric particles, the position vector $\bm{r}$ and scattering vector $\bm{q}$ can be rewritten in terms of their magnitudes $r$ and $q$, respectively, in Eq. (\ref{F_def}).
In this case, performing a substitution integral on the right-hand side of Eq. (\ref{F_def}), we obtain
\begin{equation}
    F(q) = 4 \pi \int^\infty _0 \rho(r) \frac{\sin(qr)}{qr}r^2dr .
    \label{F_2}
\end{equation}
When a spherical scatterer of radius $R_M$ with the uniform scattering length density $\rho_s$ exists in a medium with the uniform scattering length density $\rho_m$, Eq. (\ref{F_2}) can be calculated as
\begin{eqnarray}
    F(q, R_M) &=& 4\pi (\rho_s - \rho_m)\int^{R_M} _0 \frac{\sin(qr)}{qr} r^2 dr, \\
    &=& \Delta \rho V \Phi(qR_M), \label{F_result}
\end{eqnarray}
where $\Delta \rho$, $V$, and $\Phi(qR_M)$ are
\begin{equation}
    \Delta \rho = \rho_s - \rho_m, 
\end{equation}
\begin{equation}
    V = \frac{4\pi}{3}R_M^3, 
\end{equation}
\begin{equation}
    \Phi(qR_M) = \frac{3\left[\sin (qR_M) - qR_M \cos(qR_M)\right]}{(qR_M)^3}.
\end{equation}
The scattering intensity of monodisperse spheres, $I_{\text{M}}(q)$, is calculated using the number of particles per unit volume, $N_M$, as
\begin{eqnarray}
    I_{\text{M}}(q) &=& N_{\text{M}}F(q, R_M)^2, \\
    &=& \frac{\phi}{V}F(q, R_M)^2, \label{I_mono}
\end{eqnarray}
where $\phi$ is the volume fraction of one particle and $V$ is the volume of one particle. 
In the process of generating data, contributions of the background noise $b$ and measurement time $t$ are added to Eq. (\ref{I_mono}) to finally obtain the scattering intensity $I_{\text{M}}(q)$ as
\begin{equation}
    I_{\text{M}}(q) = \left(\frac{\phi}{V}F(q, R_M)^2 +b \right)\times t.
    \label{last_I_mono}
\end{equation}
\par
In general, the measurements are described by a stochastic model with noise. 
The scattering intensity is measured as the number of photons incident on the detector. 
Therefore, the noise in the measurement data is assumed to follow the Poisson distribution\cite{nagata2019} and the measured scattering intensity $y$ is given by
\begin{equation}
    y = \text{Poisson}(I_{\text{M}}(q)) \label{mono_poisson_noise}.
\end{equation}

\subsection{Polydisperse Spheres with Gaussian Size Distribution}
In this section, we formulate the scattering intensity of polydisperse spheres with Gaussian size distribution.
Using the structure factor formula for isolated monodisperse spheres Eq. (\ref{F_result}), we can express the scattering intensity $I_{\text{P}}(q)$ for polydisperse spherical particles as
\begin{equation}
    I_{\text{P}}(q) = \int^\infty _0 N_{\text{P}} F(q, r)^2 f(r) dr, \label{I_poly}
\end{equation}
where $f(r)$ is the particle size distribution function and $N_{\text{P}}$ is the number of particles per unit expressed in the following form 
\begin{equation}
    f(r) = \frac{1}{\sigma\sqrt{2\pi}} \exp \left( -\frac{(r - R_P)^2}{2\sigma^2} \right), 
\end{equation}
\begin{equation}
    N_{\text{P}} = \frac{\phi}{\langle V \rangle}.
\end{equation}
Here, $\langle V \rangle$ is the average volume of one particle, which can be written as follows using the average radius $R_P$ and standard deviation $\sigma$ of the particle size distribution:
\begin{eqnarray}
    \langle V \rangle &=& \frac{4}{3} \pi \langle r^3 \rangle, \\
    &=& \frac{4}{3} \pi R_P ^3 \left(1 + 3 \frac{\sigma^2}{R_P^2}\right).
\end{eqnarray}
In the process of generating the data, contributions of the background noise $b$ and measurement time $t$ are added to Eq. (\ref{I_poly}), and finally the scattering intensity of polydisperse spherical particles $I_{\text{P}}(q)$ is expressed as
\begin{equation}
    I_{\text{P}}(q) = \left( \int^\infty _0  \frac{\phi}{\langle V \rangle}  F(q, r)^2 f(r) dr + b\right) \times t.
    \label{last_I_poly}
\end{equation}
In the case of polydisperse spheres, as in the case of monodisperse spheres, the measurement noise is assumed to follow the Poisson distribution\cite{nagata2019} and the measured scattering intensity $y$ is given by
\begin{equation}
    y = \text{Poisson}(I_{\text{P}}(q)) \label{poly_poisson_noise}.
\end{equation}

\section{Framework}
In this section, we describe the proposed method.
In Sect. 3.1, we formulate the proposed method, and in Sects. 3.2 and 3.3, we describe the EMC method and maximum a posterior probability estimation, respectively.
 The pseudo code of the proposed method is shown in Algorithm 1.  

\subsection{Bayesian Inference}
The goal of the proposed method is to estimate the model parameters of measurement targets from the SAS data, which consist of measured magnitudes of the scattering vector and scattering intensity. 
The proposed method is based on the Bayesian inference, and the model parameters to be estimated are obtained as probability distributions. 
The reliability of estimates can be evaluated from statistics such as the peak shape and credible interval of the obtained distribution. \par
Let $\mathcal{D} = \left\{q_i, y_i\right\}_{i=1} ^{N}$ denote the measurement data and $\Theta$ denote the set of model parameters for the measurement target to be estimated. 
$N$ is the number of data. Suppose that $y_i$ is the true scattering intensity plus noise obtained in the form of Eqs. (\ref{mono_poisson_noise}) and (\ref{poly_poisson_noise}).
That is, the probability distribution of $y_i$ is the Poisson distribution \cite{nagata2019,katakami2022}:
\begin{equation}
    p(y_i|q_i, \Theta) = \frac{I(q_i;\Theta)^{y_i} \exp\left(-I(q_i;\Theta)\right)}{y_i!}.
\end{equation}
Assuming that the measured data $\mathcal{D}$ is i.i.d. under $\Theta$, the likelihood $p(\mathcal{D}|\Theta)$ can be expressed as
\begin{eqnarray}
    p(\mathcal{D}|\Theta) &=& \prod_{i=1} ^N p(y_i|q_i, \Theta), \\
    &=& \exp(-NE(\Theta)), 
\end{eqnarray}
\begin{equation}
    E(\Theta) \equiv \frac{1}{N} \sum^N _{i=1} \left\{ I(q_i;\Theta) - y_i \log I(q_i;\Theta) + \sum^{y_j}_{j=1} \log j \right\}.
    \label{poisson_E}
\end{equation}
By setting the cost function as in Eq. (\ref{poisson_E}), the model can reflect the fact that the data generation process is a Poisson point process. 
The posterior distribution of the model parameter $\Theta$, $p(\Theta | \mathcal{D})$, is obtained from Bayes' theorem using the prior distribution of $\Theta$, $p(\Theta)$, in the following form:
\begin{equation}
    p(\Theta|\mathcal{D}) = \frac{p(\mathcal{D}|\Theta)p(\Theta)}{\int p(\mathcal{D}, \Theta) d\Theta}.   \label{bayes_theorem}
\end{equation}
The proposed method samples the model parameters from the posterior distribution $p(\Theta|\mathcal{D})$.
This probability distribution is the result of estimating the distribution of the model parameters, and the shape and statistics of the obtained distribution can be used to interpret the confidence level of estimation.

\subsection{Replica Exchange Monte Carlo Method}
We use the EMC method\cite{hukushima1995} for sampling the model parameters from the posterior distribution. 
We have $L$ inverse temperatures $\beta_1,\cdots,\ \beta_L$, and consider sampling the model parameters from the joint probability distribution in the following equation:
\begin{equation}
    p(\Theta_1, \cdots, \Theta_L | \mathcal{D}, \beta_1, \cdots, \beta_L) = \prod ^{L} _{l=1} p(\Theta| \mathcal{D}, \beta_l).
\end{equation}
The posterior distribution of the parameters at the $l$-th replica $p(\Theta_l|\mathcal{D}, \beta_l)$ has the following relation from Eq. (\ref{bayes_theorem}):
\begin{equation}
    p(\Theta_l|\mathcal{D}, \beta_l) \propto \exp\left( -N \beta_l E(\Theta_l) \right) p(\Theta_l).
    \label{post}
\end{equation}
In this study, the inverse temperature is set such that  $0=\beta_1 < \cdots < \beta_L = 1$ . 
 The target probability distribution in this case is the posterior distribution at the replica with the lowest temperature where $\beta_L = 1$. 
The EMC method samples the probability distribution using the metropolis method\cite{mcmc,metropolis} for each of the $L$  Markov chains, which are called replicas with different inverse temperatures as described above, and exchanges the parameters of neighboring replicas with the probability of satisfying the detailed balance condition at any point in time.
The probability $W(\Theta_l|\Theta_{l+1})$ of exchanging the $l$-th and $l+1$-th replicas is expressed as
\begin{eqnarray}
    W(\Theta_l|\Theta_{l+1}) &=& \text{min} \left[1,\ \frac{p(\Theta_{l+1}|\mathcal{D}, \beta_{l}) p(\Theta_{l}|\mathcal{D}, \beta_{l+1})}{p(\Theta_{l}|\mathcal{D}, \beta_{l}) p(\Theta_{l+1}|\mathcal{D}, \beta_{l+1})} \right], \\
    &=& \text{min} \left[1,\ \exp \left(N(\beta_{l+1} - \beta_l)(E(\Theta_{l+1}) - E(\Theta_{l})) \right) \right].
\end{eqnarray}
Even if the solution is trapped in a local solution in the low-temperature replica, the EMC method is expected to enable the sampling of the globally optimal solution using the high-temperature replica by replica exchange.

\subsection{Maximum A Posterior Estimation}
The proposed method estimates the distribution of model parameters from the SAS data, and the point estimates correspond to the solution for the maximum a posterior probability (MAP) estimation.
The formulation of the MAP solution is given below.
Consider the case where Eq. (\ref{post}) is sampled using the EMC method described in Sect. 3.2.
Let us assume that the posterior distribution at the $L$-th replica corresponds to the target distribution.
The MAP solution is the set of model parameters that maximizes the posterior probability among the samples of model parameters for the $L$-th replica, written as $\Theta_\text{MAP}$. Using Eq. (\ref{post}), we express $\Theta_\text{MAP}$ as
\begin{equation}
    \Theta_\text{MAP} = \argmax_{\Theta_L} \exp\left( -N \beta_L E(\Theta_L) \right) p(\Theta_L).
\end{equation}

\begin{figure}[!t]
    \begin{algorithm}[H]
        \caption{Bayesian Estimation of the Measurement Target Parameters}
        \label{alg1}
        \begin{algorithmic}[1]
        \REQUIRE The measurement data, $\mathcal{D} = \{q_i, y_i\}_{i=1} ^N$. The number of replicas, $L$. The inverse tempretures, $\{\beta_l\}_{l=1}^L$ where $0 = \beta_1 < \cdots < \beta_L = 1$.
        The Burn-In, $S_0$. The number of samples, $S_1$. The step size for the Metropolis algorithm, $\Delta s$. The prior distribution of parameters, $p(\Theta)$.
        \ENSURE $\forall l$ $\in \{1, \cdots, L\}$, $\forall s \in \{1, \cdots,S_1\} $, $\Theta_l^s \sim p(\Theta_l|\mathcal{D}, \beta_l)$.
        \STATE Initialize array of sampled parameters, $\Psi = \{\}$.
        \FOR{$l \in \{1, \cdots, L \}$}
        \STATE $\Theta_l^0 \sim p(\Theta)$
        \ENDFOR
        \FOR{$s \in \{1, \cdots, S_0 + S_1\}$}
        \FOR{$l \in \{1, \cdots, L \}$}
        \STATE Propose the following state, $\Theta'$ = $\Theta_l^{s-1}$ + $\Delta s \times \text{Uniform}(-1, 1)$.
        \STATE Calculate the acceptance ratio, $\alpha = {p(\Theta'|\mathcal{D}, \beta_l)}/{p(\Theta_l^{s-1}|\mathcal{D}, \beta_l)}$.
        \IF{$\text{Uniform}(0, 1) < \alpha$}
        \STATE $\Theta_l^{s}$ = $\Theta'$
        \ELSE
        \STATE $\Theta_l^{s}$ = $\Theta_l^{s-1}$
        \ENDIF
        \ENDFOR
        \FOR{$l \in \{1, \cdots ,L-1 \}$}
        \STATE Calculate the probability of exchanging states, $W(\Theta_l^s|\Theta_{l+1}^s)$.
        \IF{$\text{Uniform}(0, 1) < W(\Theta_l^s|\Theta_{l+1}^s)$}
        \STATE Swap the $\Theta_l^s$ for the $\Theta_{l+1}^s$.
        \ENDIF
        \ENDFOR
        \IF{$s > S_0$}
        \STATE Append the $\{\Theta_l^s\}_{l=1}^L$ to the $\Psi$.
        \ENDIF
        \ENDFOR
        \end{algorithmic}
    \end{algorithm}
\end{figure}

\newpage

\section{Numerical Experiments}
In this section, we present the procedure and results of numerical experiments using artificial data.
In these numerical experiments, the model parameters are estimated using the proposed method presented in Sect. 3 from the artificial data generated using the models for the two types of measurement target formulated in Sect. 2.
The process of generating artificial data is as follows.
First, $N$ values of the magnitude of the scattering vector $q$ are given in the interval $[q_{\text{min}}, q_{\text{max}}]$ at equal intervals.
For each of the $q$ values, we calculate the measured scattering intensity $y$ using Eq. (\ref{mono_poisson_noise}) and Eq. (\ref{poly_poisson_noise}). 
The artificial data $\{q_i, y_i \}_{i=1} ^N$ are obtained by the above procedure.
Data points where the measured scattering intensity $y$ is zero due to Poisson noise are used in the simulation but are omitted when plotting.\par
This section deals with numerical experiments that vary the lower limit of the angular region of the measurement data $q_\text{min}$ and measurement time $t$, as well as numerical experiments with extremely small number of data $N$.  
The reason for this is to find the estimation limit of the proposed method for the lower limit of the magnitude of the scattering vector $q_\text{min}$ of the measurement data, since it is difficult to obtain measurement data in small-angle regions in actual experiments. 
In addition, since it is desirable for the measurement time $t$ to be small to improve the efficiency of the measurement, the estimation limit of the proposed method for the measurement time $t$ is also obtained.
Furthermore, we describe experiments with extremely small number of data in order to discuss how the estimation results change as the number of data decreases.

In the following Sects. 4.1 and 4.2, we describe the setup and results of numerical experiments for the monodisperse and polydisperse spheres models, respectively.
In the experiments in Sects. 4.1 and 4.2, $N = 400$ was used. 
In Sect. 4.3, numerical experiments where the number of non-zero values in the data is 10 are described.   \par
The programs for these numerical experiments were developed in C++ and parallelized using the OpenMP API. 
The longest-running case took about 3 hours to complete on an Apple M1 Pro chip with 10 cores and 32 GB of RAM.

\subsection{Experiments Using Monodisperse Sphere Model}
In this section, we describe the setup and results of numerical experiments on the monodisperse spheres model.
Table I shows the values of the parameters for the model expressed by Eq. (\ref{last_I_mono}) used to generate artificial data. 
The parameters to be estimated for the monodisperse spheres model are the particle radius $R_M$, background noise $b$, and measurement time $t$, where $\Theta = \{R_M, b, t \}$. 
The true values of model parameters are shown in Table I and denoted as $\Theta^* = \{R_M ^*, b^*, t^* \}$.
The prior distribution of the model parameter $p(\Theta)$ to be estimated is defined as follows. 
The parameters of the prior distribution given by Eq. (\ref{mono_pori_1})-(\ref{monp_pri_2}) were determined empirically.
The initial values of $\Theta$ are given by random numbers that follow the prior distribution of each parameter.
\begin{equation}
    p(\Theta) = p(R_M)p(b)p(t),
\end{equation}
\begin{eqnarray}
    p(R_M) &=& \text{Gamma}(R_M; \alpha_R, \gamma_R), \label{monp_pri_1} \\
    &=& \frac{e^{-x/\gamma_R}}{\gamma_R ^{\alpha_R} \cdot \Gamma(\alpha_R)} \cdot x^{\alpha_R - 1},
\end{eqnarray}
\begin{equation}
    p(b) = \text{Gamma}(b; \alpha_b, \gamma_b),
    \label{mono_b}
\end{equation}
\begin{equation}
    p(t) = \text{Gamma}(t; \alpha_t, \gamma_t).
    \label{mono_t}
\end{equation}
\begin{eqnarray}
    \alpha_R &=& 1.5,\ \gamma_R = 100, \label{mono_pori_1}\\
    \alpha_b &=& 1.8,\ \gamma_b = 1, \\
    \alpha_t &=& 1.1,\ \gamma_t = 500. \label{monp_pri_2}
\end{eqnarray}
\noindent In the numerical experiments on the artificial data generated using the monodisperse spheres model, the number of replicas for the EMC method is $L=40$ and the inverse temperatures $\beta_l\ (l=1, \cdots, L)$ of each replica are defined as
\begin{equation}
    \beta_l = \left\{ 
        \begin{array}{ll}
            0 & (\text{if\ \ } l =1)\\
            2.2^{l-L} & (\text{otherwise}).
        \end{array}
    \right.
\end{equation}

\begin{table}[h]
    \centering
    \caption{Parameter values for generating artificial data using monodisperse spheres model.}
    \begin{tabular}{|r||r|}
    \hline
    Radius $R_M^*$ (nm) & 10 \\
    \hline
    Background noise $b^*$ (cm $^{-1}$) & 0.01 \\
    \hline
    Measurement time $t^*$ & 10, 1, 0.1 \\
    \hline
    Volume fraction $\phi$ & 1 \\
    \hline
    Scattering length density of solvent, $\rho_s$ (nm$^{-2}$) & 1e-4 \\
    \hline
    Scattering length density of medium, $\rho_m$ (nm$^{-2}$) & 6.3e-4 \\
    \hline
    \end{tabular}
\end{table}

In the following, we present the results of numerical experiments in which we analyzed the artificial data generated with the monodisperse spheres model using the proposed method. 
 The results presented in this section were obtained by taking the burn-in of the EMC method as $10^5$ and subsequently sampling $10^5$ times. \par
First, we set $q_{\text{min}} = 0.01\ \text{nm}^{-1}$ and $q_{\text{max}} = 3.0\ \text{nm}^{-1}$, and describe the results for three types of artificial data generated with measurement time $t^*=10,\ 1,\ 0.1$.
The plots of the artificial data are shown in Fig. 2. Figure 3 shows the results of fitting using the MAP solution. 
Figure 4 shows histograms of the posterior distributions of the sampled model parameters.
From Fig. 3, it can be seen that as the measurement time $t^*$ decreases, the region where no data are obtained with the intensity less than 0 increases; thus, the accuracy of the fitting of the lower part of the scattering function becomes unstable.
This is due to the fact that the peak position of the distribution of the background noise $b$, which contributes to the vertical position of the scattering intensity shown in Fig. 4, deviates from the true value more significantly than those of the distributions of other parameters as the measurement time $t^*$ decreases.
The distributions of the $R_M$ and $t$ shown in Fig. 4 widen as $t^*$ decreases, indicating that the estimation uncertainty increases.

\begin{figure}[H]
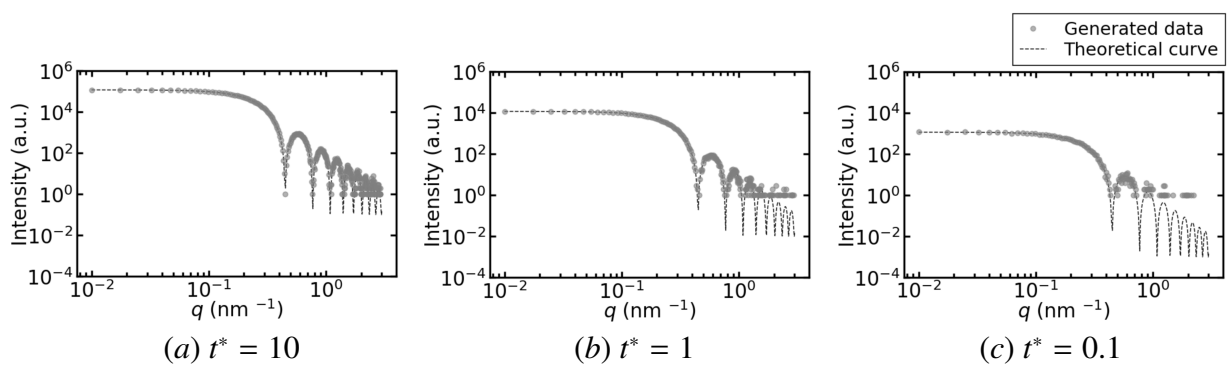

    \begin{minipage}[b]{0.335\linewidth}
        \hspace{-6mm}
        \includegraphics[keepaspectratio, scale=0.22]{70773_Fig2_a.pdf}
        \vspace{-4mm}
        \begin{center}
            \hspace{-3mm}$(a)\ t^*=10$
        \end{center}
    \end{minipage} 
    \begin{minipage}[b]{0.35\linewidth}
        \hspace{-5mm}
        \includegraphics[keepaspectratio, scale=0.22]{70773_Fig2_b.pdf}
        \vspace{-4mm}
        \begin{center}
        \hspace{-6mm}$(b)\ t^*=1$
        \end{center}
    \end{minipage}
    \begin{minipage}[b]{0.3\linewidth}
        \hspace{-6mm}
        \includegraphics[keepaspectratio, scale=0.22]{70773_Fig2_c.pdf}
        \vspace{-11mm}
        \begin{center}
            $(c)\ t^*=0.1$
        \end{center}
    \end{minipage}

    \caption{Plots of artificial data generated using the monodisperse spheres model with $t^*=10,\ 1$, and $0.1$ under the settings $q_{\text{min}}=0.01\ \text{nm}^{-1}$ and $q_{\text{max}} = 3.0 \text{nm}^{-1}$.
    The gray dots represent the measured data and the black dashed line represents the true scattering intensity. }
\end{figure}

\begin{figure}[H]
    \begin{minipage}[b]{0.33\linewidth}
        \hspace{-6mm}
        \includegraphics[keepaspectratio, scale=0.22]{70773_Fig3_a.pdf}
        \vspace{-4mm}
        \begin{center}
         $(a)\ t^*=10$
        \end{center}
    \end{minipage} 
    \begin{minipage}[b]{0.35\linewidth}
        \hspace{-5mm}
        \includegraphics[keepaspectratio, scale=0.22]{70773_Fig3_b.pdf}
        \vspace{-4mm}
        \begin{center}
        \hspace{-3mm} $(b)\ t^*=1$
        \end{center}
    \end{minipage}
    \begin{minipage}[b]{0.3\linewidth}
        \hspace{-7mm}
        \includegraphics[keepaspectratio, scale=0.22]{70773_Fig3_c.pdf}
        \vspace{-4mm}
        \begin{center}
            $(c)\ t^*=0.1$
        \end{center}
    \end{minipage}

    \caption{(Color online) Results of fitting the monodisperse spheres model using the MAP solution to artificial data when $t^*=10,\ 1$ and $0.1$ under the settings $q_\text{min}=0.01\ \text{nm}^{-1}$ and $q_{\text{max}} = 3.0\ \text{nm}^{-1}$. 
    The gray dots are the measured data, the black dashed line is the true scattering intensity, and the red line is the scattering intensity obtained using the MAP solution. }
\end{figure}

\begin{figure}[H]
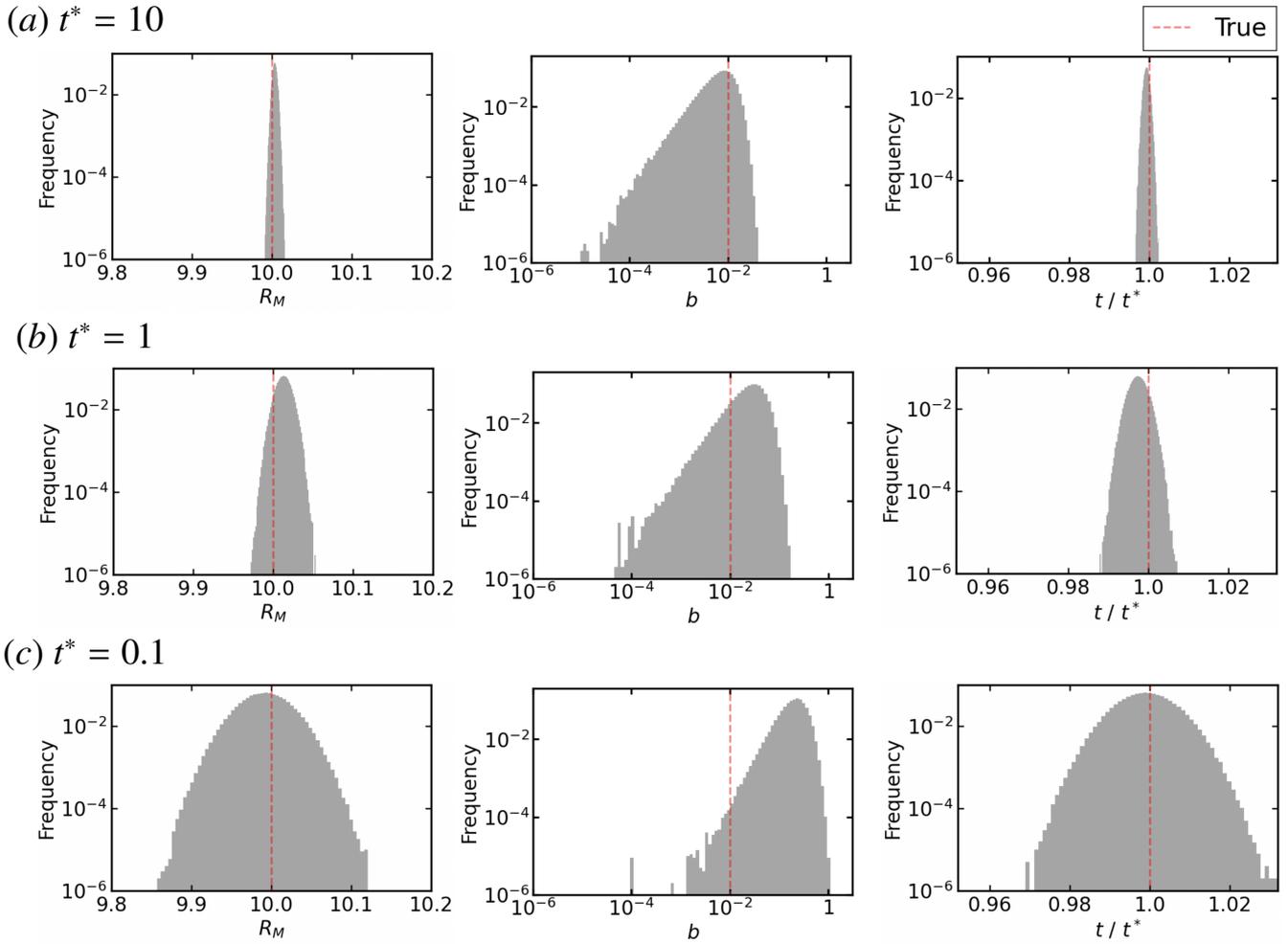

    \hspace{-15cm} 
    \begin{tabular}{ccc}
        \begin{minipage}[b]{0.35\linewidth}
            \begin{center}
                \large $\hspace{-6cm} (a)\ t^*=10$
            \end{center}
            \hspace*{-1cm}
            \includegraphics[keepaspectratio, scale=0.23]{70773_Fig4_a_r.pdf}
        \end{minipage}
        \begin{minipage}[b]{0.35\linewidth}
            \hspace*{-0.8cm}
            \includegraphics[keepaspectratio, scale=0.23]{70773_Fig4_a_b.pdf}
        \end{minipage}
        \begin{minipage}[b]{0.3\linewidth}
            \hspace*{-0.5cm}
            \includegraphics[keepaspectratio, scale=0.23]{70773_Fig4_a_t.pdf}
        \end{minipage} \\
        
        \begin{minipage}[b]{0.35\linewidth}
            \begin{center}
                \large $\hspace{-6cm} (b)\ t^*=1$
            \end{center}
            \hspace*{-1cm}
            \includegraphics[keepaspectratio, scale=0.23]{70773_Fig4_b_r.pdf}
        \end{minipage}
        \begin{minipage}[b]{0.35\linewidth}
            \hspace*{-0.8cm}
            \includegraphics[keepaspectratio, scale=0.23]{70773_Fig4_b_b.pdf}
        \end{minipage}
        \begin{minipage}[b]{0.3\linewidth}
            \hspace*{-0.5cm}
            \includegraphics[keepaspectratio, scale=0.23]{70773_Fig4_b_t.pdf}
        \end{minipage} \\
        
        \begin{minipage}[b]{0.35\linewidth}
            \begin{center}
                \large $\hspace{-6cm} (c)\ t^*=0.1$
            \end{center}
            \hspace*{-1cm}
            \includegraphics[keepaspectratio, scale=0.23]{70773_Fig4_c_r.pdf}
        \end{minipage}
        \begin{minipage}[b]{0.35\linewidth}
            \hspace*{-0.8cm}
            \includegraphics[keepaspectratio, scale=0.23]{70773_Fig4_c_b.pdf}
        \end{minipage}
        \begin{minipage}[b]{0.3\linewidth}
            \hspace*{-0.5cm}
            \includegraphics[keepaspectratio, scale=0.23]{70773_Fig4_c_t.pdf}
        \end{minipage}
    \end{tabular}
     \caption{(Color online) Histograms of posterior distributions of parameters sampled using the artificial data generated using the monodisperse spheres model when $t^*=10,\ 1$ and $0.1$ under the settings $q_{\text{min}}=0.01\ \text{nm}^{-1}$ and $q_{\text{max}} = 3.0\ \text{nm}^{-1}$.
     The red dashed line represents the true value. Histograms of $t$ are expressed as $t / t^*$. Here the number of bins is set to 64. }
  \end{figure}

Next, we show the results for three types of artificial data generated at $q_\text{min} = 0.4,\ 2.35$ and $2.65\ \text{nm}^{-1}$ under the settings $q_\text{max} = 3.0\ \text{nm}^{-1}$ and $t^*=10$.
These settings correspond to the case where data for small-angle region could not be obtained in the actual experiment.
The plots of the artificial data are shown in Fig. 5, and the results of fitting using the MAP solution are shown in Fig. 6. 
Histograms of the posterior distributions of the sampled model parameters are shown in Fig. 7.
Figures 3 and 6 show that the accuracy of the fitting hardly changes with respect to the $q_\text{min}$ of the data. 
Figure 7 shows that as the $q_\text{min}$ of the measured data increases, the shape of the distribution of $b$ hardly changes, but the width of the distribution of $t$ increases. 
The distribution of $R_M$ in Fig. 6(c) has multiple peaks, indicating that the proposed method has reached its estimation limit for $q_\text{min}$ at the time of that setup. \par
Residual plots, corresponding to the fittings shown in Figs. 3 and 6, are shown in Figs. A$\cdot$1 and A$\cdot$2 in the Appendix, respectively. 
Table I\hspace{-1.2pt}I shows the estimation results of the parameters obtained from the numerical experiments using the monodisperse spheres model described in this section.
The values in the table are those of $\Theta_\text{MAP} {}^{+p}_{-q}$, where $p$ is the difference between the MAP solution $\Theta_\text{MAP}$ and the upper limit of the 99\% credible interval and $q$ is the difference between $\Theta_\text{MAP}$ and the lower limit for each estimated model parameter.

\begin{figure}[H]
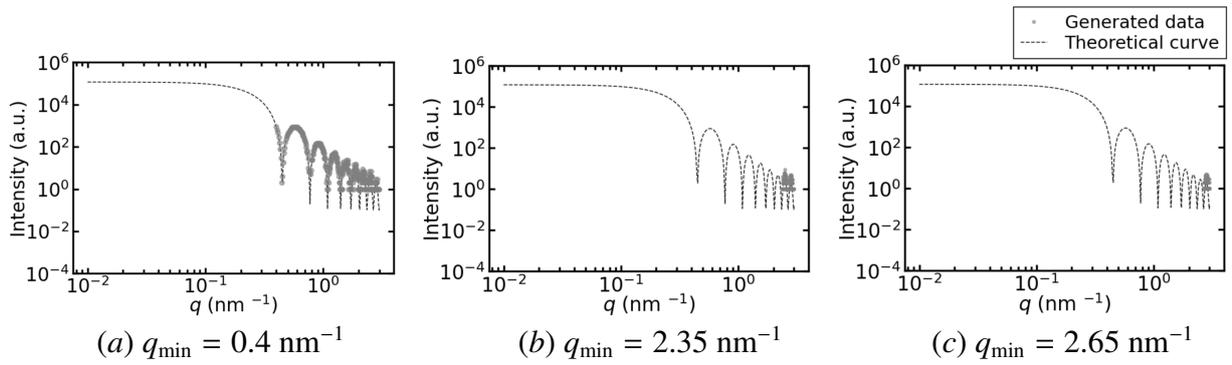

    \centering
    \begin{minipage}[b]{0.335\linewidth}
        \hspace{-6mm}
        \includegraphics[keepaspectratio, scale=0.21]{70773_Fig5_a.pdf}
        \vspace{-4mm}
        \begin{center}
        \hspace{-7mm} $(a)\ q_\text{min} = 0.4\ \text{nm}^{-1}$
        \end{center}
    \end{minipage}
    \begin{minipage}[b]{0.35\linewidth}
        \hspace{-5mm}
        \includegraphics[keepaspectratio, scale=0.22]{70773_Fig5_b.pdf}
        \vspace{-4mm}
        \begin{center}
            \hspace{-3mm} $(b)\ q_\text{min} = 2.35\ \text{nm}^{-1}$
        \end{center}
    \end{minipage}
    \begin{minipage}[b]{0.3\linewidth}
        \hspace{-6mm}
        \includegraphics[keepaspectratio, scale=0.22]{70773_Fig5_c.pdf}
        \vspace{-11mm}
        \begin{center}
            \hspace{2mm} $(c)\ q_\text{min} = 2.65\ \text{nm}^{-1}$
        \end{center}
    \end{minipage}

    \caption{Plot of artificial data generated using the monodisperse spheres model when $q_\text{min}=0.4,\ 2.35$ and $2.65\ \text{nm}^{-1}$ under the settings $q_\text{max} = 3.0\ \text{nm}^{-1}$ and $t^*=10$.
            The gray dots represent the measured data and the black dashed line represents the true scattering intensity. }
\end{figure}

\begin{figure}[H]
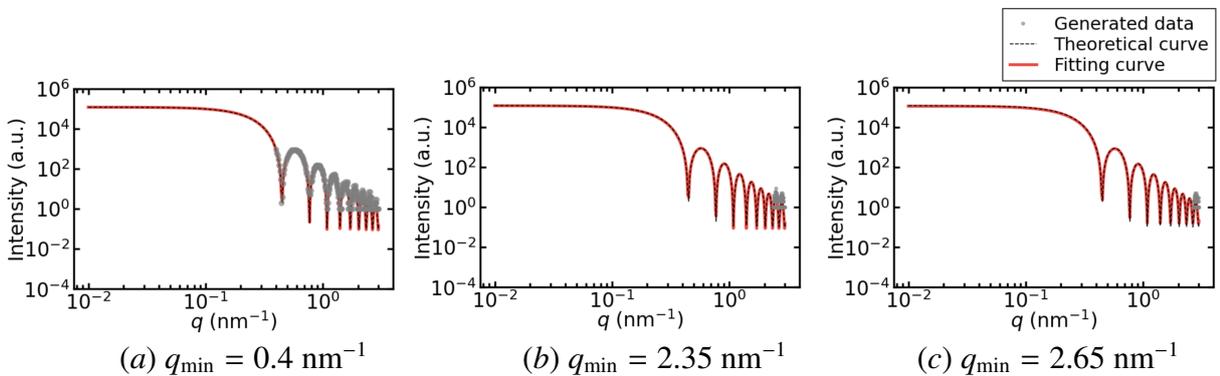

    \centering
    \begin{minipage}[b]{0.33\linewidth}
        \hspace{-7mm}
        \includegraphics[keepaspectratio, scale=0.22]{70773_Fig6_a.pdf}
        \vspace{-4mm}
        \begin{center}
            $(a)\ q_\text{min} = 0.4\ \text{nm}^{-1}$
        \end{center}
    \end{minipage}
    \begin{minipage}[b]{0.35\linewidth}
        \hspace{-6mm}
        \includegraphics[keepaspectratio, scale=0.22]{70773_Fig6_b.pdf}
        \vspace{-4mm}
        \begin{center}
            $(b)\ q_\text{min} = 2.35\ \text{nm}^{-1}$
        \end{center}
    \end{minipage}
    \begin{minipage}[b]{0.3\linewidth}
        \hspace{-7mm}
        \includegraphics[keepaspectratio, scale=0.22]{70773_Fig6_c.pdf}
        \vspace{-4mm}
        \begin{center}
            $(c)\ q_\text{min} = 2.65\ \text{nm}^{-1}$
        \end{center}
    \end{minipage}

    \caption{(Color online) Results of fitting the monodisperse spheres model using the MAP solution to artificial data when $q_\text{min}=0.4,\ 2.35$ and $2.65\ \text{nm}^{-1}$ under the setting $q_\text{max} = 3.0\ \text{nm}^{-1}$ and $t^*=10$. 
            The gray dots are the measured data, the black dashed line is the true scattering intensity, and the red line is the scattering intensity calculated using the MAP solution. }
\end{figure}

\begin{figure}[H]
    \begin{tabular}{ccc}
        \begin{minipage}[b]{0.35\linewidth}
            \begin{center}
                \large $\hspace{-5cm} (a)\ q_\text{min}=0.4$
            \end{center}
            \hspace*{-1cm}
            \includegraphics[keepaspectratio, scale=0.23]{70773_Fig7_a_r.pdf}
        \end{minipage}
        \begin{minipage}[b]{0.35\linewidth}
            \hspace*{-0.8cm}
            \includegraphics[keepaspectratio, scale=0.23]{70773_Fig7_a_b.pdf}
        \end{minipage}
        \begin{minipage}[b]{0.3\linewidth}
            \hspace*{-0.5cm}
            \includegraphics[keepaspectratio, scale=0.23]{70773_Fig7_a_t.pdf}
        \end{minipage} \\
        
        \begin{minipage}[b]{0.35\linewidth}
            \begin{center}
                \large $\hspace{-5cm} (b)\ q_\text{min}=2.35$
            \end{center}
            \hspace*{-1cm}
            \includegraphics[keepaspectratio, scale=0.23]{70773_Fig7_b_r.pdf}
        \end{minipage}
        \begin{minipage}[b]{0.35\linewidth}
            \hspace*{-0.8cm}
            \includegraphics[keepaspectratio, scale=0.23]{70773_Fig7_b_b.pdf}
        \end{minipage}
        \begin{minipage}[b]{0.3\linewidth}
            \hspace*{-0.5cm}
            \includegraphics[keepaspectratio, scale=0.23]{70773_Fig7_b_t.pdf}
        \end{minipage} \\
        
        \begin{minipage}[b]{0.35\linewidth}
            \begin{center}
                \large $\hspace{-5cm} (c)\ q_\text{min}=2.65$
            \end{center}
            \hspace*{-1cm}
            \includegraphics[keepaspectratio, scale=0.23]{70773_Fig7_c_r.pdf}
        \end{minipage}
        \begin{minipage}[b]{0.35\linewidth}
            \hspace*{-0.8cm}
            \includegraphics[keepaspectratio, scale=0.23]{70773_Fig7_c_b.pdf}
        \end{minipage}
        \begin{minipage}[b]{0.3\linewidth}
            \hspace*{-0.5cm}
            \includegraphics[keepaspectratio, scale=0.23]{70773_Fig7_c_t.pdf}
        \end{minipage}
    \end{tabular}
    \caption{(Color online) Histograms of the posterior distributions of parameters sampled using the artificial data generated using the monodisperse spheres model when $q_\text{min}=0.4,\ 2.35$ and $2.65\ \text{nm}^{-1}$ under the settings $q_\text{max} = 3.0\ \text{nm}^{-1}$ and $t^*=10$.
            The red dashed line represents the true value. Histograms of $t$ are expressed as $t/t^*$.   Here the number of bins is set to 64. }
\end{figure}

\begin{table}[H]
    \centering
    \caption{Estimation results of the parameters obtained from the numerical experiments using the monodisperse spheres model.}
    \begin{tabular}{|c|c||c|c|c|} 
        \hline
        $q_\text{min} \ [\text{nm}^{-1}]$ & $t^*$ & $R_M \ [\text{nm}]$ & $b \ [\text{cm}^{-1}]$ & $t$\\ \hline \hline

        0.01 & 10 & $10.00 ^{+0.007}_{-0.007}$ & $(5.53 ^{+16.93} _{-5.17}) \times 10^{-3}$ & $9.99 ^{+0.02} _{-0.02}$ \\ \cline{2-5}
             & 1 & $10.01 ^{+0.02}_{-0.02}$ & $(1.96 ^{+6.82}_{-1.80}) \times 10^{-2}$ & $1.00^{+0.01} _{-0.01}$  \\ \cline{2-5}
             & 0.1 &$9.99 ^{+0.08} _{-0.07}$ & $(1.70 ^{+4.28} _{-1.44}) \times 10^{-1}$ & $(1.00 ^{+0.02}_{-0.02}) \times 10^{-1}$ \\ \cline{1-5}
        \hline
            0.4 & 10 & $10.00 ^{+0.01} _{-0.01}$ & $(8.71 ^{+15.10}_{-7.01}) \times 10^{-3}$ & $10.06 ^{+0.15}_{-0.16}$ \\
        \hline
            2.35 & 10 & $10.02 ^{+0.03} _{-0.03}$ & $(8.74^{+15.03}_{-6.15}) \times 10^{-3}$ & $9.85 ^{+1.25} _{-1.55}$ \\
        \hline
            2.65 & 10 & $10.01 ^{+1.18}_{-1.16}$ & $(1.60 ^{+1.49}_{-1.01}) \times 10^{-2}$ & $9.51 ^{+4.05} _{-2.06}$ \\ \cline{1-5}
    \end{tabular}
\end{table}

\subsection{Experiments Using Polydisperse Sphere Model}
In this section, we describe the setup and results of numerical experiments using artificial data generated using polydisperse spheres with Gaussian size distribution. 
Table I\hspace{-1.2pt}I\hspace{-1.2pt}I shows the parameter values of the model expressed in Eq. (\ref{last_I_poly}) that were used to generate the artificial data. 
The parameters to be estimated for the polydisperse spheres model are the mean radius of particles, $R_P$, the standard deviation of the radius distribution, $\sigma$, the background noise $b$, and the measurement time $t$, where $\Theta = \{R_P, \sigma, b, t \}$.
The true values of each model parameter are those shown in Table I\hspace{-1.2pt}I\hspace{-1.2pt}I and denoted as $\Theta^* = \{R_P^*, \sigma^*, b^*, t^* \}$. 
The prior distribution of $R_P$ is defined as follows, using the same form as the prior distribution of the radius $R_M$ for the monodisperse spheres model given by Eq. (\ref{monp_pri_1}):
\begin{equation}
    p(R_P) = \text{Gamma}(R_P; \alpha_{R}, \gamma_{R}).
\end{equation}
The prior distribution of $\sigma$ is defined as follows. 
\begin{equation}
    p(\sigma) = \text{Gamma}(\sigma; \alpha_\sigma, \beta_\sigma),
\end{equation}
\begin{eqnarray}
    \alpha_\sigma = 1.8,\ \beta_\sigma = 50. \label{sig_prior}
\end{eqnarray}
The parameters of the prior distribution of $\sigma$ given by Eq. (\ref{sig_prior}), are determined empirically.
The prior distributions of $b$ and $t$ are given by Eq. (\ref{mono_b}) and (\ref{mono_t}), which are equal to the distributions used for the monodisperse spheres model.
The prior distribution $p(\Theta)$ of the model parameter $\Theta$ has the following form:
\begin{equation}
    p(\Theta) = p(R_P)p(\sigma)p(b)p(t).
\end{equation}
The initial values of $\Theta$ were given by random numbers following prior distributions of each parameter.
In numerical experiments using artificial data of the polydisperse spheres model, the number of replicas for the EMC method is $L=32$ and the inverse temperatures $\beta_l\ (l=1, \cdots, L)$ of each replica are defined as
\begin{equation}
    \beta_l = \left\{ 
        \begin{array}{ll}
            0 & (\text{if\ \ } l =1)\\
            1.7^{l-L} & (\text{otherwise}).
        \end{array}
    \right. 
\end{equation}

\begin{table}[H]
    \centering
    \caption{Parameter values for generating artificial data using polydisperse spheres model.}
    \begin{tabular}{|r||r|}
    \hline
    Mean radius $R_P^*$ (nm) & 10 \\
    \hline
    Standard deviation $\sigma^*$ & 2 \\
    \hline
    Background noise $b^*$ (cm $^{-1}$) & 0.001 \\
    \hline
    Measurement time $t^*$ & 100, 10, 1 \\
    \hline
    Volume fraction $\phi$ & 0.01 \\
    \hline
    Scattering length density of solvent, $\rho_s$ (nm$^{-2}$) & 1e-4 \\
    \hline
    Scattering length density of medium, $\rho_m$ (nm$^{-2}$) & 6.3e-4 \\
    \hline
    \end{tabular}
\end{table}

In the following, we present the results of numerical experiments in which we analyzed artificial data generated using the polydisperse spheres model by the proposed method. 
The results presented in this section were obtained by taking the burn-in of the EMC method as $10^5$ and subsequently sampling $10^5$ times. \par
First, we describe the results for three types of artificial data generated with $t^*=100, 10$, and $1$ under the settings $q_\text{min} = 0.01\ \text{nm}^{-1}$ and $q_\text{max} = 7.0\ \text{nm}^{-1}$.
The plots of the artificial data are shown in Fig. 8. 
Figure 9 shows the results of fitting these artificial data using the MAP solution. 
Figure 10 shows histograms of the posterior distribution of the sampled model parameters.
The fitting results in Fig. 9 show that, as is the case for the monodisperse spheres model data, as the measurement time decreased, the number of regions with the intensity less than 0 where no data are obtained increases; thus, the fitting of the lower part of the scattering function becomes unstable.
As shown in Fig. 10, the distribution corresponding to each model parameter becomes wider as $t^*$ decreases, and the estimation uncertainty increases. 
Histograms of the posterior distribution shown in Fig. 10(c) are very wide, indicating that estimation becomes more difficult. 
The reason for this is considered to be that, as shown in Fig. 8(c), the data at the measurement time $t^*=1$ does not include the shoulder and lower part information that is characteristic of the true scattering intensity function.

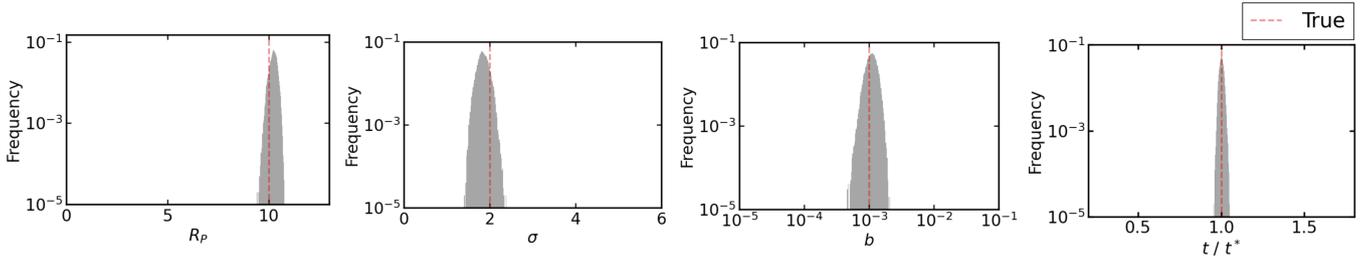
\begin{figure}[H]
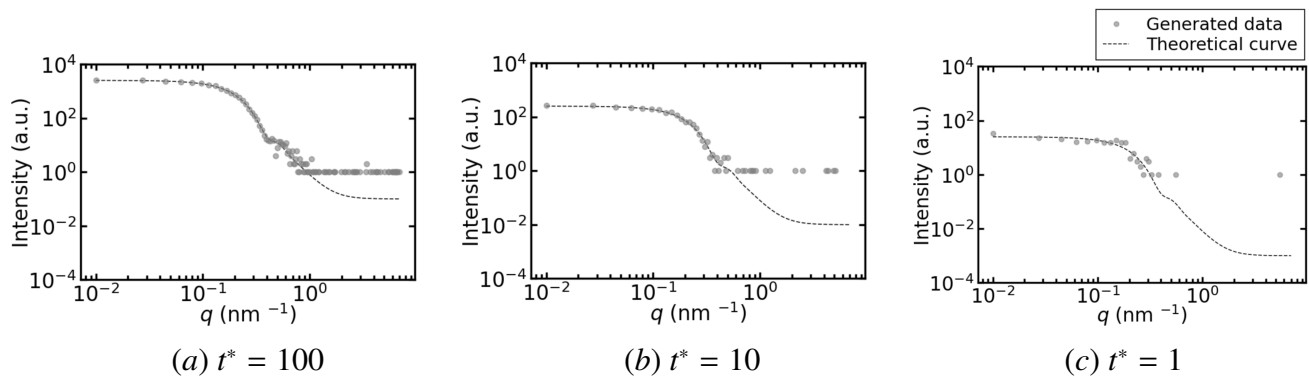

    \centering
    \begin{minipage}[b]{0.335\linewidth}
        \hspace{-11mm}
        \includegraphics[keepaspectratio, scale=0.23]{70773_Fig8_a.pdf}
        \vspace{-3mm}
        \begin{center}
            \hspace{-10mm} $(a)\ t^*=100$
        \end{center}
    \end{minipage}
    \begin{minipage}[b]{0.35\linewidth}
        \hspace{-5mm}
        \includegraphics[keepaspectratio, scale=0.23]{70773_Fig8_b.pdf}
        \vspace{-3mm}
        \begin{center}
            \hspace{-1mm}$(b)\ t^*=10$
        \end{center}
    \end{minipage}
    \begin{minipage}[b]{0.3\linewidth}
        \hspace{-5mm}
        \includegraphics[keepaspectratio, scale=0.215]{70773_Fig8_c.pdf}
        \vspace{-10mm}
        \begin{center}
        \hspace{8mm} $(c)\ t^*=1$
        \end{center}
    \end{minipage}

    \caption{Plot of artificial data for the polydisperse spheres model with $t^*=100, 10$, and $1$ under the settings $q_\text{min}=0.01\ \text{nm}^{-1}$ and $q_\text{max} = 7.0\ \text{nm}^{-1}$.
            The gray dots represent artificial data and the black dashed line represents the true scattering intensity. }
\end{figure}

\begin{figure}[H]
    \centering
    \begin{minipage}[b]{0.33\linewidth}
        \hspace{-11mm}
        \includegraphics[keepaspectratio, scale=0.23]{70773_Fig9_a.pdf}
        \vspace{-3mm}
        \begin{center}
            \hspace{-6mm}$ (a)\ t^*=100$
        \end{center}
    \end{minipage} 
    \begin{minipage}[b]{0.35\linewidth}
        \hspace{-5mm}
        \includegraphics[keepaspectratio, scale=0.23]{70773_Fig9_b.pdf}
        \vspace{-3mm}
        \begin{center}
            $(b)\ t^*=10$
        \end{center}
    \end{minipage}
    \begin{minipage}[b]{0.3\linewidth}
        \hspace{-6mm}
        \includegraphics[keepaspectratio, scale=0.23]{70773_Fig9_c.pdf}
        \vspace{-10mm}
        \begin{center}
            \hspace{12mm}$(c)\ t^*=1$
        \end{center}
    \end{minipage}

    \caption{(Color online) Results of fitting the polydisperse spheres model using the MAP solution for artificial data when $t^*=100, 10$, and $1$ under the settings $q_\text{min}=0.01\ \text{nm}^{-1}$ and $q_\text{max} = 7.0\ \text{nm}^{-1}$.
            The gray dots are the measured data, the black dashed line is the true scattering intensity, and the red line is the scattering intensity calculated using the MAP solution. }
\end{figure}

\begin{figure}[H]
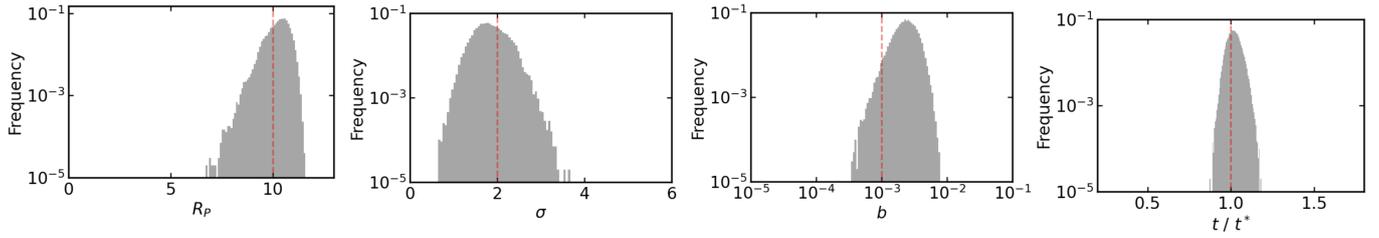
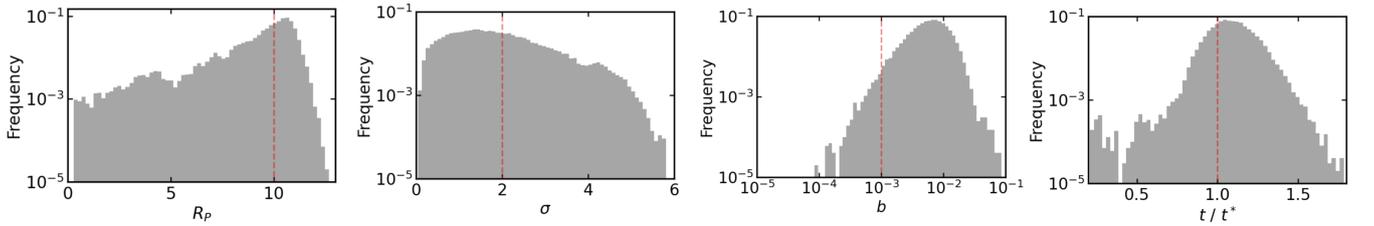

    \begin{tabular}{ccc}
        \begin{minipage}[b]{1.4\linewidth}
            \begin{center}
                \large $\hspace{-22.5cm} (a)\ t^*=100$
            \end{center}
            \hspace*{-1.5cm}
            \includegraphics[keepaspectratio, scale=0.3]{70773_Fig10_a.pdf}
        \end{minipage} \\
        
        \begin{minipage}[b]{1.4\linewidth}
            \begin{center}
                \large $\hspace{-22.8cm} (b)\ t^*=10$
            \end{center}
            \hspace*{-1.6cm}
            \includegraphics[keepaspectratio, scale=0.305]{70773_Fig10_b.pdf}
        \end{minipage} \\
        
        \begin{minipage}[b]{1.4\linewidth}
            \begin{center}
                \large $\hspace{-22.8cm} (c)\ t^*=1$
            \end{center}
            \hspace*{-1.6cm}
            \includegraphics[keepaspectratio, scale=0.3]{70773_Fig10_c.pdf}
        \end{minipage}
    \end{tabular}

    \caption{(Color online) Histograms of the posterior distribution of each model parameter sampled using the artificial data of the polydisperse spheres model when $t^*=100, 10$, and $1$ under the settings $q_\text{min}=0.01\ \text{nm}^{-1}$ and $q_\text{max} = 7.0\ \text{nm}^{-1}$. 
            The red dashed line represents the true value. Histograms for $t$ are expressed as $t / t^*$. Here the number of bins is set to 64. }
  \end{figure}

Next, we show the results for two types of artificial data generated with $q_\text{min}=0.2$ and $0.3\ \text{nm}^{-1}$ under the settings $q_\text{max}=7.0$ and $t^*=100$.
These settings correspond to the case where data for the small-angle region could not be obtained in the actual experiment.
Figure 11 shows the artificial data in this case, and Fig. 12 shows the results of fitting using the MAP solution. 
Histograms of the posterior distributions of the sampled model parameters are shown in Fig. 13. 
The fitting results shown in Figs. 9 and 12 indicate that the accuracy of the fitting hardly changes with $q_\text{min}$.
In Fig. 13(b), the estimation limit for $q_\text{min}$ is reached, the local solution begins to be sampled, and histograms of the distribution for model parameters have multiple peaks.
\par
Residual plots, corresponding to the fittings shown in Figs. 9 and 12, are shown in Figs. A$\cdot$3 and A$\cdot$4 in the Appendix, respectively.
Table I\hspace{-1.2pt}V shows the estimation results of the model parameters for the numerical experiments on the polydisperse spheres model treated in this section. 
The values in the table are those of $\Theta_\text{MAP} {}^{+p}_{-q}$, where $p$ is the difference between the MAP solution $\Theta_\text{MAP}$ and the upper limit of the 99\% credible interval and $q$ is the difference between $\Theta_\text{MAP}$ and the lower limit for each estimated model parameter.

  \begin{figure}[H]
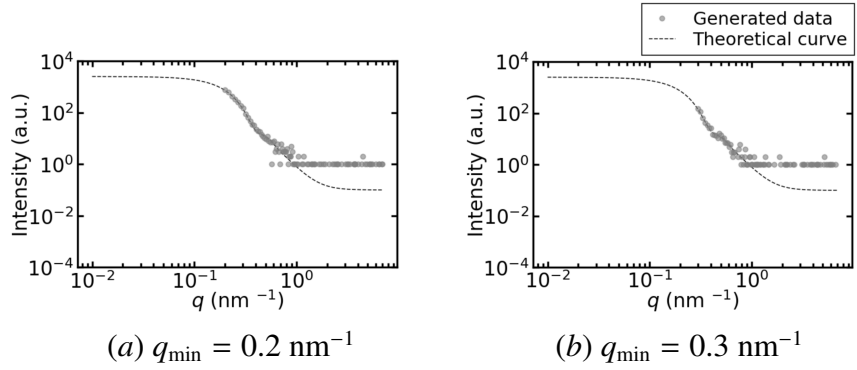

    \centering
    \begin{minipage}[b]{0.335\linewidth}
        \hspace{-8mm}
        \includegraphics[keepaspectratio, scale=0.22]{70773_Fig11_a.pdf}
        \vspace{-3mm}
        \begin{center}
        \hspace{-8mm} $(a)\ q_\text{min}=0.2\ \text{nm}^{-1}$
        \end{center}
    \end{minipage} 
    \begin{minipage}[b]{0.35\linewidth}
        \includegraphics[keepaspectratio, scale=0.22]{70773_Fig11_b.pdf}
        \vspace{-3mm}
        \begin{center}
        \hspace{1mm} $(b)\ q_\text{min}=0.3\ \text{nm}^{-1}$
        \end{center}
    \end{minipage} 

    \caption{Plot of artificial data for the polydisperse spheres model when $q_\text{min}=0.2$ and $0.3\ \text{nm}^{-1}$ under the settings $q_\text{max}=7.0$ and $t^*=100$. 
            The gray dots are the measured data, and the black dashed line is the true scattering intensity.}
\end{figure}

\begin{figure}[H]
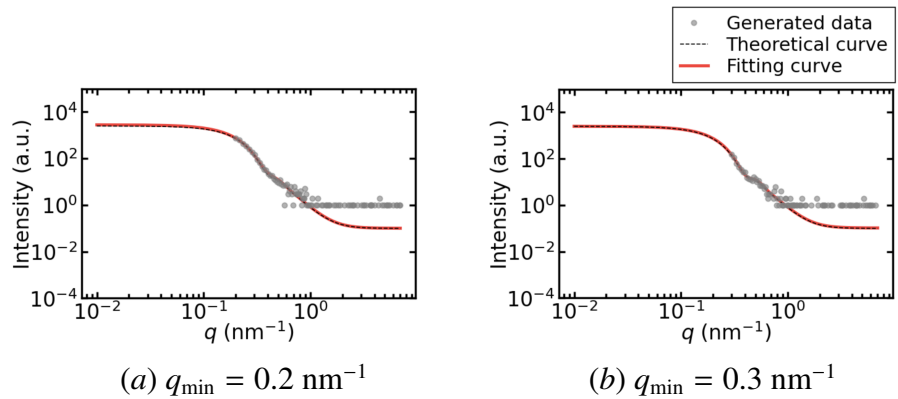

    \centering
    \begin{minipage}[b]{0.35\linewidth}
        \hspace{-8mm}
        \includegraphics[keepaspectratio, scale=0.23]{70773_Fig12_a.pdf}
        \vspace{-3mm}
        \begin{center}
            \hspace{-8mm} $(a)\ q_\text{min}=0.2\ \text{nm}^{-1}$
        \end{center}
    \end{minipage}
    \begin{minipage}[b]{0.35\linewidth}
        \includegraphics[keepaspectratio, scale=0.23]{70773_Fig12_b.pdf}
        \vspace{-10mm}
        \begin{center}
            \hspace{4mm} $(b)\ q_\text{min}=0.3\ \text{nm}^{-1}$
        \end{center}
    \end{minipage}

    \caption{(Color online) Results of fitting the polydisperse spheres model using the MAP solution to artificial data when $q_\text{min}=0.2$ and $0.3\ \text{nm}^{-1}$ under the settings $q_\text{max}=7.0$ and $t^*=100$. 
            The gray dots are the measured data, the black dashed line is the true scattering intensity, and the red line is the scattering intensity calculated from the MAP solution. }
\end{figure}

\begin{figure}[H]
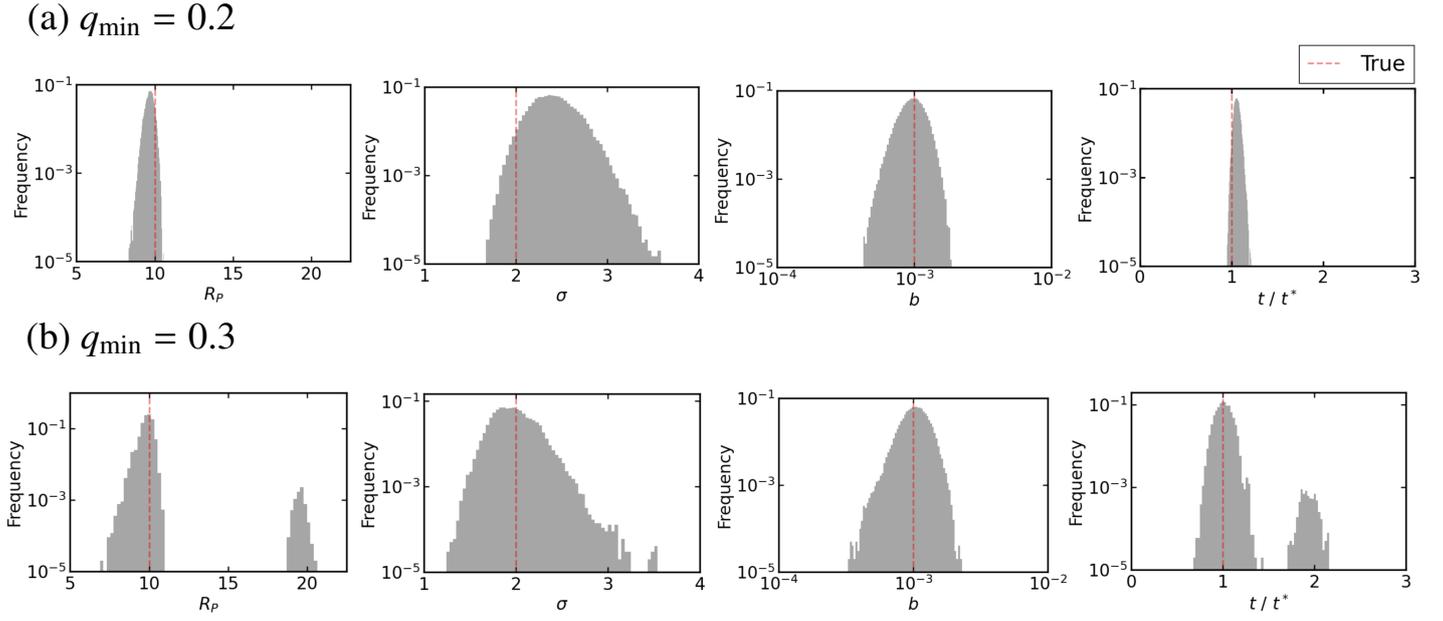

    \begin{tabular}{ccc}
        \begin{minipage}[b]{1.4\linewidth}
            \begin{center}
                \large \hspace{-22.5cm} (a) $q_\text{min}=0.2$
            \end{center}
            \vspace{-1.8mm}
            \hspace{-2cm}
            \includegraphics[keepaspectratio, scale=0.3]{70773_Fig13_a.pdf}
        \end{minipage} \\
        
        \begin{minipage}[b]{1.4\linewidth}
            \begin{center}
                \large \hspace{-22.5cm} (b) $q_\text{min}=0.3$
            \end{center}
            \vspace{2mm}
            \hspace{-2cm}
            \includegraphics[keepaspectratio, scale=0.31]{70773_Fig13_b.pdf}
        \end{minipage} \\
    \end{tabular}

     \caption{(Color online) Histograms of the posterior distribution of each model parameter sampled using the artificial data of the polydisperse spheres model when $q_\text{min}=0.2$ and $0.3\ \text{nm}^{-1}$ under the settings $q_\text{max}=7.0$ and $t^*=100$. 
              The red dashed line represents the true value. Histograms for $t$ represent the $t / t^*$ values. Here the number of bins is set to 64. }
\end{figure}

\begin{table}[H]
    \centering
    \caption{Estimation results of the parameters obtained from the numerical experiments using the polydisperse spheres model.}
    \begin{tabular}{|c|c||c|c|c|c|} 
        \hline
        $q_\text{min} \ [\text{nm}^{-1}]$ & $t^*$ & $R_P \ [\text{nm}]$ & $\sigma$ & $b \ [\text{cm}^{-1}]$ & $t$\\ \hline \hline
        0.01 & 100 & $10.25 ^{+0.36} _{-0.48}$ & $1.83 ^{+0.32} _{-0.29}$ & $(1.07 ^{+0.59}_{-0.39}) \times 10^{-3}$ & $100.25 ^{+3.19}_{-2.97}$ \\ \cline{2-6}
             & 10 & $10.43 ^{+0.78}_{-1.93}$ & $1.75 ^{+1.09}_{-0.72}$ & $(2.13 ^{+3.25}_{-1.42}) \times 10^{-3}$ & $10.16 ^{+1.08}_{-0.88}$ \\ \cline{2-6}
             & 1 & $10.44 ^{+1.17} _{-9.10}$ & $1.28 ^{+3.67} _{-1.10}$ & $(4.42 ^{+20.31}_{-3.73}) \times 10^{-3}$ & $1.08 ^{+0.39}_{-0.36}$ \\ \cline{1-6}
        0.2 & 100 & $9.72 ^{+0.43} _{-0.57}$ & $2.35 ^{+0.51}_{-0.34}$ & $(1.00 ^{+0.41}_{-0.29}) \times 10^{-3}$ & $105.50 ^{+6.99}_{-4.70}$ \\
        \hline
        0.3 & 100 & $10.01 ^{+9.43}_{-1.51}$ & $1.90 ^{+0.72} _{-0.39}$ & $(1.00 ^{+0.70}_{-0.46}) \times 10^{-3}$ & $103.75 ^{+82.24}_{-20.66}$ \\
        \hline
    \end{tabular}
\end{table}

\subsection{Experiments with Extremely Small Number of Data}
In the analysis with our proposed method, as the number of non-zero data points decreases, the width of the posterior distribution corresponding to the estimation uncertainty increases accordingly. 
In this section, we describe numerical experiments using measured data with 10 non-zero data points for the monodisperse and polydisperse spheres models respectively, 
in order to discuss the change in confidence of the estimates as the number of data points decreases.
The results presented in this section were obtained by taking the burn-in of the EMC method as $10^5$ and subsequently sampling $10^5$ times.\par
First, we consider the case of the monodisperse spheres model and describe the setup and results of the numerical experiment on artificial data with measurement time of $t^* = 10$. 
In the numerical experiment with artificial data for the monodisperse spheres model, the number of data $N$ is set to 11 to ensure that the number of non-zero data points is 10.
The artificial data were generated with $q_\text{min}=0.01\ \text{nm}^{-1}$ and $q_\text{max} = 3.0\ \text{nm}^{-1}$. 
The parameters used to generate artificial data, the prior distribution, and the forward model are the same as those used in Sect. 4.1.
The number of replicas is $L = 40$ and the inverse temperatures $\beta_l\ (l = 1, \cdots , L)$ of each replica are determined using the following equation,
\begin{equation}
    \beta_l = \left\{ 
        \begin{array}{ll}
            0 & (\text{if\ \ } l =1)\\
            2.1^{l-L} & (\text{otherwise}).
        \end{array}
    \right. 
\end{equation}

Plot of the artificial data is shown in Fig. 14. 
The model parameters $\Theta = \{R_M , b, t\}$ are estimated from the data, and the result of fitting using the MAP solution is shown in Fig. 15.
The histogram of the posterior distribution sampled is shown in Fig. 16.
Figure 16 demonstrates that while the peak of the distribution for the background noise deviates from the true value, the estimation of the distribution for the radius and measurement time is accurate. \par
A residual plot, corresponding to the fitting shown in Fig. 15, is shown in Fig. A$\cdot$5 in the Appendix.
Table V summarizes the estimation results of the model parameters for the numerical experiments on the monodisperse spheres model treated in this section. 
The values in the table represent the values of $\Theta_\text{MAP} {}^{+p}_{-q}$, where the $p$ and the $q$ are the difference between the MAP solution $\Theta_\text{MAP}$ and the upper and lower limits of the 99\% credible interval for each estimated model parameter.

\begin{figure}[H]
    \centering
    \begin{minipage}[b]{0.3\linewidth}
        \centering
        \includegraphics[keepaspectratio, scale=0.22]{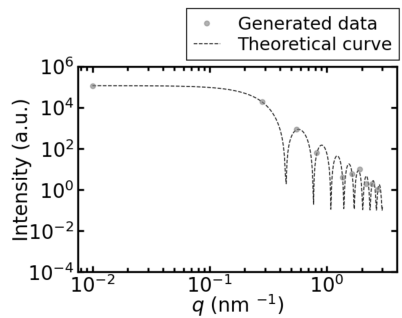}
    \end{minipage}

    \caption{Plot of artificial data for the monodisperse spheres model with 10 non-zero data.
            The gray dots are the measured data, the black dashed line is the true scattering intensity. }
\end{figure}

\begin{figure}[H]
    \centering
    \begin{minipage}[b]{0.3\linewidth}
        \centering
        \includegraphics[keepaspectratio, scale=0.22]{70773_Fig15.pdf}
    \end{minipage}

    \caption{(Color online) Result of fitting the monodisperse spheres model using the MAP solution to artificial data with 10 non-zero data.
             The gray dots are the measured data, the black dashed line is the true scattering intensity, and the red line is the scattering intensity calculated from the MAP solution.  }
\end{figure}

\begin{figure}[H]
    \begin{tabular}{ccc}
        \begin{minipage}[b]{0.30\linewidth}
            \centering
            \includegraphics[keepaspectratio, scale=0.2]{70773_Fig16_r.pdf}
        \end{minipage}
        \begin{minipage}[b]{0.35\linewidth}
            \centering
            \includegraphics[keepaspectratio, scale=0.2]{70773_Fig16_b.pdf}
        \end{minipage}
        \begin{minipage}[b]{0.3\linewidth}
            \centering
            \includegraphics[keepaspectratio, scale=0.2]{70773_Fig16_t.pdf}
        \end{minipage}
    \end{tabular}
     \caption{(Color online) Histograms of the posterior distribution of each model parameter sampled using the artificial data of the monodisperse spheres model with 10 non-zero data.
              The red dashed line represents the true value. The histogram for the $t$ represent the values $t / t^*$.Here the number of bins is set to 64. }
  \end{figure}

  \begin{table}[H]
    \centering
    \caption{ Estimation result for the artificial data of the monodisperse spheres model with 10 non-zero data. }
    \begin{tabular}{|c||c|c|c|c|} 
        \hline
        $t^*$ & $R_M \ [\text{nm}]$ & $b \ [\text{cm}^{-1}]$ & $t$\\ \hline \hline
        10 & $10.00 ^{+0.20}_{-0.04}$ & $(4.28 ^{+4.10} _{-4.10}) \times 10^{-2}$ & $9.99 ^{+0.12} _{-0.12}$ \\ 
        \hline
    \end{tabular}
\end{table}

Next, we consider the case of the polydisperse spheres model. 
Using the same parameters for artificial data generation, the prior distribution, and the forward model as in Sect. 4.2, 
we describe the setup and results of numerical experiments for artificial data with measurement time of $t^* = 100$. 
In the numerical experiment with artificial data for the polydisperse spheres model, the number of data $N$ is set to 42 to ensure that the number of non-zero data points is 10.
The artificial data were generated with $q_\text{min}=0.01\ \text{nm}^{-1}$ and $q_\text{max} = 7.0\ \text{nm}^{-1}$. 
The number of replicas is $L = 32$, and the inverse temperatures $\beta_l\ (l = 1, \cdots, L)$ of each replica are determined using the following equation, 
\begin{equation}
    \beta_l = \left\{ 
        \begin{array}{ll}
            0 & (\text{if\ \ } l =1)\\
            1.69^{l-L} & (\text{otherwise}).
        \end{array}
    \right. 
\end{equation}

Plot of the artificial data is shown in Fig. 17. 
The model parameters $\Theta = \{R_P, \sigma, b, t \}$ are estimated from the data, 
and the fitting results using the MAP solution are shown in Fig. 18.
The histogram of the posterior distribution sampled is shown in Fig. 19. 
From Fig. 19, we can see that in all distributions, the peaks are located around the true values, but there are areas where the shape is distorted.\par
A residual plot, corresponding to the fitting shown in Fig. 18, is shown in Figs. A$\cdot$6 in the Appendix. 
Table V\hspace{-1.2pt}I summarizes the estimation results of the model parameters for the numerical experiments on the polydisperse spheres model treated in this section. 
The values in the table represent the values of $\Theta_\text{MAP} {}^{+p}_{-q}$, where the $p$ and the $q$ are the difference between the MAP solution $\Theta_\text{MAP}$ 
and the upper and lower limits of the 99\% credible interval for each estimated model parameter.

\begin{figure}[H]
    \centering
    \begin{minipage}[b]{0.3\linewidth}
        \centering
        \includegraphics[keepaspectratio, scale=0.22]{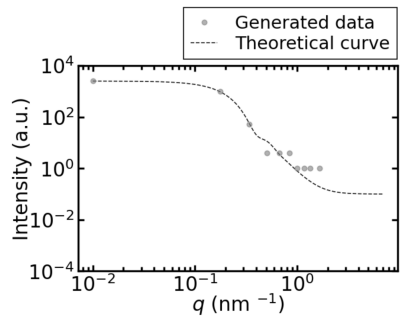}
    \end{minipage}

    \caption{ Plot of artificial data for the polydisperse spheres model with 10 non-zero data.
    The gray dots are the measured data, the black dashed line is the true scattering intensity. }
\end{figure}

\begin{figure}[H]
    \centering
    \begin{minipage}[b]{0.3\linewidth}
        \centering
        \includegraphics[keepaspectratio, scale=0.22]{70773_Fig18.pdf}
    \end{minipage}

    \caption{(Color online) Result of fitting the polydisperse spheres model using the MAP solution to artificial data with 10 non-zero data.
    The gray dots are the measured data, the black dashed line is the true scattering intensity, and the red line is the scattering intensity calculated from the MAP solution. }
\end{figure}

\begin{figure}[H]
    \vspace{-0.5cm}
    \hspace{-1.3cm}
    \includegraphics[keepaspectratio, scale=0.35]{70773_Fig19.pdf}
    \caption{(Color online) Histograms of the posterior distribution of each model parameter sampled using the artificial data of the polydisperse spheres model with 10 non-zero data.
     The red dashed line represents the true value. The histogram for the $t$ represent the values $t / t^*$.Here the number of bins is set to 64. }
\end{figure}

\begin{table}[H]
    \centering
    \caption{ Estimation result for the artificial data of the polydisperse spheres model with 10 non-zero data. }
    \begin{tabular}{|c||c|c|c|c|c|} 
        \hline
        $t^*$ & $R_P \ [\text{nm}]$ & $\sigma$ & $b \ [\text{cm}^{-1}]$ & $t$\\ \hline \hline
        100 & $10.17 ^{+0.72} _{-1.93}$ & $1.84 ^{+0.77} _{-0.63}$ & $(7.60 ^{+22.65}_{-3.48}) \times 10^{-4}$ & $102.29 ^{+8.63}_{-9.37}$ \\
        \hline
    \end{tabular}
\end{table}

\section{Discussion}
In this section, we discuss the results of the numerical experiments presented in Sect. 4. 
 In that section, we presented the estimation results of model parameters using the proposed method for artificial data of monodisperse and polydisperse spheres models across various measurement times and angular regions, as well as numerical experiments with an extremely small number of data for each model.   
In this study, measurement noise is assumed to follow Poisson distribution, so the longer the measurement time $t^*$, the lower the noise intensity.
This is why the results presented in Sect. 4 show that when the range of the magnitudes of the scattering vector $q$ of the measured data is equal, the histogram of the posterior distribution sampled for data with large values of the measurement time $t^*$ has a MAP solution closer to the true value and spreads less than when $t^*$ is small.
As shown in Figs. 2 and 8, the smaller the value of the measurement time $t^*$, the harder it is to obtain data for the lower part of the true scattering intensity function.
Therefore, as the measurement time $t^*$ decreases, the peak of the distribution for the background noise $b$, which contributes to the vertical direction of the scattering intensity function, shifts away from the true value.
On the other hand, Figs. 3, 6, 9, and 12 show that the accuracy of the fitting hardly changes with the lower limit for the magnitude of the scattering vector $q_\text{min}$.
Tables I\hspace{-1.2pt}I and I\hspace{-1.2pt}V show that the MAP solution hardly changes with $q_\text{min}$. 
These results indicate that the lower limit for the magnitude of the scattering vector $q_\text{min}$ in measured data has little effect on the value of the MAP solution estimated by the proposed method in the experimental setup described in Sect. 4.
\par
We now discuss the estimation limits of the proposed method for the measurement time $t^*$ and the lower bound of the scattering vector $q_\text{min}$.
First, we discuss the monodisperse spheres model treated in Sect. 4.1.
Figure 4 and Table I\hspace{-1.2pt}I show that the value of the MAP solution for the background noise $b$ at $t^*=0.1$ deviates from the true value by more than one order of magnitude and that the 99\% credible interval of the posterior distribution does not include the true value.
This suggests that the transition point of estimability for the measurement time $t^*$ is in the range of $t^*=0.1 \-- 1$ in the case of the monodisperse spheres model.
Figure 7(c) shows that histograms of the posterior distribution of $R_M$ for data with $q_\text{min} = 2.65\ \text{nm}^{-1}$ have multiple peaks. 
This suggests that the transition point of estimability for the lower limit of the angle region $q_\text{min}$ is in the range of $q_\text{min}=2.35 \-- 2.65\ \text{nm}^{-1}$ in the case of the monodisperse spheres model. 
Next, we discuss the polydisperse spheres model presented in Sect. 4.2. 
Figure 10(c) shows that histograms of the posterior distribution of each parameter are wider than those in Fig. 10(a) and 10(b), and the distribution for the measurement time $t$ splits into two peaks. 
This suggests that the transition point of estimability for the measurement time $t^*$ is in the range of $t^*=1 \-- 10$ in the case of the polydisperse spheres model. 
On the other hand, Fig. 13(b) shows that histograms of the posterior distributions of $R_P$ and $t$ have two peaks. 
Therefore, in the case of the polydisperse spheres model, the transition point of estimability for the lower limit of the angle region $q_\text{min}$ is considered to be in the range $q_\text{min}=0.2 \-- 0.3\ \text{nm}^{-1}$. \par
As discussed above, our proposed method enables us to analyze data with short measurement time and missing small angle regions, and to interpret the confidence level of estimation.
As shown in Tables I\hspace{-1.2pt}I and I\hspace{-1.2pt}V, if the measurement time and the lower bounds of the angular region are above the transition point of estimability, even for artificial data with short measurement time or limited angular region, the 99\% credible interval of the posterior distribution for the model parameters contains the true value.
In addition, the distributions of $R_M$ in Fig. 7(c) and those of $R_P$, $\sigma$, and $t$ in Fig. 13(b), which are considered to be in the estimation limit, have multiple peaks, but the MAP solution is very close to the true value, as shown in Tables I\hspace{-1.2pt}I and I\hspace{-1.2pt}V.\par
  
In section 4.3, the results of parameter estimation were presented for the case where the number of data points was extremely small where the number of non-zero data is 10.
Comparing the histograms of the posterior distributions obtained in the two experiments with the monodisperse spheres model, where only the number of data differs, Figures 4(a) and 16 show that the histograms in Fig. 16 are wider in all cases. 
This is also the case with the polydisperse spheres model. 
Comparing the histograms of the posterior distributions obtained from two experiments with only a difference in the number of data points, it can be seen that the distribution shown in Fig. 19 is wider and more distorted in shape than all the other distributions in Fig. 10(a).
These may indicate an increase in the estimation uncertainty due to the small number of data.

\section{Conclusions}
In this paper, we propose a method of estimating distributions of model parameters of measurement targets from the measured data obtained in SAS experiments using the Bayesian inference.
Numerical experiments on artificial data were conducted to evaluate the performance of the proposed method. 
From the numerical experiments, we confirmed that the estimation results of the proposed method indicate the reliability of estimates and the transition point of estimability for the measurement time and the lower bounds of the angular region. 
 Furthermore, we confirmed that parameter estimation is possible even when the number of data is extremely small, and that changes in the confidence level of the estimation with respect to the number of data can be interpreted in terms of the width and shape of the distribution.  \par
In the numerical experiments, model parameters such as those related to size were estimated from artificial data generated using monodisperse and polydisperse spheres with Gaussian size distribution, which are representative measurement target models.
However, analysis targets of the proposed method are not limited to these models and parameters. 
It is possible to analyze a measurement target using various models and parameters. \par
 
In our proposed analysis, a model of measurement targets is defined for the measured data, and the model parameters are estimated using a Bayesian framework.
This paper deals with cases where the models corresponding to the measurement data are known, but in actual analysis, there may be several candidate models, and it may be necessary to evaluate the validity of each and choose one.
This problem is called model selection.
In addition, while this study dealt with cases where the background noise of the measurement data was constant, real data often have complex background noise, and it may be necessary to consider a new modeling method for such cases. 
These problems and the analysis of publicly available data, such as McSAS\cite{McSAS} data, and real-world data are challenges to be tackled in the future.

\begin{acknowledgement}
    This work was supported by JST CREST (Grant Nos. PMJCR1761 and JPMJCR1861) from the Japan Science and Technology Agency (JST).
\end{acknowledgement}

\appendix
\section{Residual Plot}
Residual plots for the fitting results shown in Figs. 3, 6, 9, 12, 15, and 18 are shown in Figs. A$\cdot$1 -- A$\cdot$6, respectively. 
Here, the residuals for data points with zero scattering intensity are omitted, and the residuals are normalized by the estimated scattering intensity values.
The horizontal axis of these residual plots is represented by the logarithmic axis.
Considering the nature of Poisson noise, which was considered as measurement noise in this study, data with low scattering intensity contain large noise. 
This makes fitting more difficult for the high-angle data, which is more affected by noise than the low-angle data. 
This is consistent with the residual plots in Figs. A$\cdot$1 -- A$\cdot$6.

\begin{figure}[H]
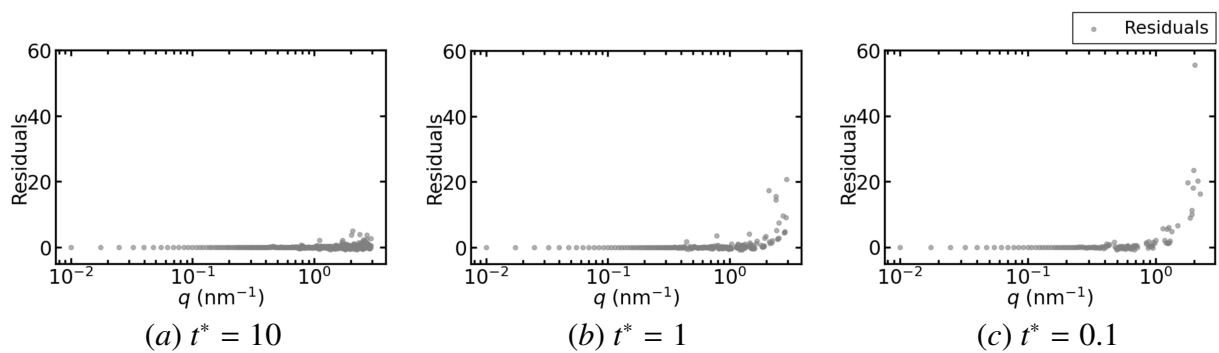

    \centering
    \begin{minipage}[b]{0.33\linewidth}
        \hspace{-5mm}
        \includegraphics[keepaspectratio, scale=0.22]{70773_FigA1_a.pdf}
        \vspace{-4mm}
        \begin{center}
            \hspace{-6mm}$ (a)\ t^*=10$
        \end{center}
    \end{minipage} 
    \begin{minipage}[b]{0.35\linewidth}
        \hspace{-3mm}
        \includegraphics[keepaspectratio, scale=0.22]{70773_FigA1_b.pdf}
        \vspace{-4mm}
        \begin{center}
            \hspace{-5mm}$(b)\ t^*=1$
        \end{center}
    \end{minipage}
    \begin{minipage}[b]{0.3\linewidth}
        \hspace{-5mm}
        \includegraphics[keepaspectratio, scale=0.22]{70773_FigA1_c.pdf}
        \vspace{-11mm}
        \begin{center}
            \hspace{2mm}$(c)\ t^*=0.1$
        \end{center}
    \end{minipage}
    \caption{Residual plots of the fitting to the monodisperse spheres data with $t^*=10, 1, 0.1$ under the settings $q_\text{min}=0.01\ \text{nm}^{-1}$ and $q_{\text{max}} = 3.0\ \text{nm}^{-1}$.}
\end{figure}

\begin{figure}[H]
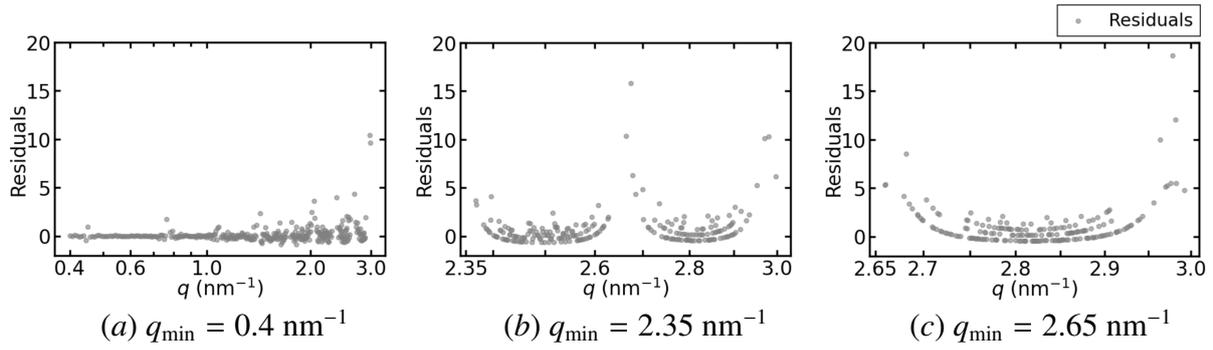

    \centering
    \begin{minipage}[b]{0.33\linewidth}
        \hspace{-4mm}
        \includegraphics[keepaspectratio, scale=0.22]{70773_FigA2_a.pdf}
        \vspace{-4mm}
        \begin{center}
            $(a)\ q_\text{min} = 0.4\ \text{nm}^{-1}$
        \end{center}
    \end{minipage}
    \begin{minipage}[b]{0.35\linewidth}
        \hspace{-3mm}
        \includegraphics[keepaspectratio, scale=0.22]{70773_FigA2_b.pdf}
        \vspace{-4mm}
        \begin{center}
            $(b)\ q_\text{min} = 2.35\ \text{nm}^{-1}$
        \end{center}
    \end{minipage}
    \begin{minipage}[b]{0.3\linewidth}
        \hspace{-5mm}
        \includegraphics[keepaspectratio, scale=0.22]{70773_FigA2_c.pdf}
        \vspace{-11mm}
        \begin{center}
        \hspace{2mm} $(c)\ q_\text{min} = 2.65\ \text{nm}^{-1}$
        \end{center}
    \end{minipage}
    \caption{Residual plots of the fitting to the monodisperse spheres data when $q_\text{min}=0.4,\ 2.35,\ 2.65\ \text{nm}^{-1}$ under the settings $q_\text{max} = 3.0\ \text{nm}^{-1}$ and $t^*=10$. }
\end{figure}

\begin{figure}[H]
    \centering
    \begin{minipage}[b]{0.33\linewidth}
        \hspace{-5mm}
        \includegraphics[keepaspectratio, scale=0.22]{70773_FigA3_a.pdf}
        \vspace{-4mm}
        \begin{center}
            \hspace{-6mm}$ (a)\ t^*=100$
        \end{center}
    \end{minipage} 
    \begin{minipage}[b]{0.35\linewidth}
        \hspace{-4mm}
        \includegraphics[keepaspectratio, scale=0.22]{70773_FigA3_b.pdf}
        \vspace{-4mm}
        \begin{center}
            \hspace{-8mm}$(b)\ t^*=10$
        \end{center}
    \end{minipage}
    \begin{minipage}[b]{0.3\linewidth}
        \hspace{-6mm}
        \includegraphics[keepaspectratio, scale=0.22]{70773_FigA3_c.pdf}
        \vspace{-11mm}
        \begin{center}
            \hspace{-5mm}$(c)\ t^*=1$
        \end{center}
    \end{minipage}
    \caption{Residual plots of the fitting to the polydisperse spheres data with $t^*=100, 10, 1$ under the settings $q_\text{min}=0.01\ \text{nm}^{-1}$ and $q_\text{max} = 7.0\ \text{nm}^{-1}$. }
\end{figure}

\begin{figure}[H]
    \centering
    \begin{minipage}[b]{0.35\linewidth}
        \hspace{-8mm}
        \includegraphics[keepaspectratio, scale=0.22]{70773_FigA4_a.pdf}
        \vspace{-4mm}
        \begin{center}
            \hspace{-8mm} $(a)\ q_\text{min}=0.2\ \text{nm}^{-1}$
        \end{center}
    \end{minipage}
    \begin{minipage}[b]{0.35\linewidth}
        \hspace{-1mm}
        \includegraphics[keepaspectratio, scale=0.22]{70773_FigA4_b.pdf}
        \vspace{-4mm}
        \begin{center}
            \hspace{4mm} $(b)\ q_\text{min}=0.3\ \text{nm}^{-1}$
        \end{center}
    \end{minipage}

    \caption{Residual plots of the fitting to the polydisperse spheres data when $q_\text{min}=0.2,\ 0.3\ \text{nm}^{-1}$ under the settings $q_\text{max}=7.0\ \text{nm}^{-1}$ and $t^*=100$.}
\end{figure}

\begin{figure}[H]
    \centering
    \begin{minipage}[b]{0.35\linewidth}
        \centering
        \hspace{-0.7cm}
        \includegraphics[keepaspectratio, scale=0.23]{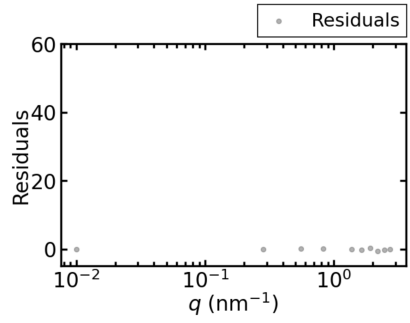}
    \end{minipage}
    \vspace{-4mm}
    \caption{Residual plot of the fitting to the monodisperse spheres data with 10 non-zero data($N=11$).}
\end{figure}

\begin{figure}[H]
    \centering
    \begin{minipage}[b]{0.35\linewidth}
        \centering
        \hspace{-0.7cm} 
        \includegraphics[keepaspectratio, scale=0.23]{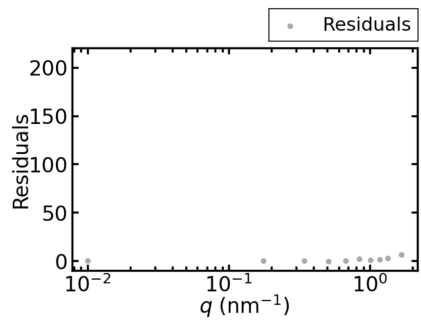}
    \end{minipage}
    \vspace{-4mm}
    \caption{Residual plot of the fitting to the polydisperse spheres data with 10 non-zero data($N=42$).}
\end{figure}

\end{document}